# Effect of Spatial Charge Inhomogeneity on 1/f Noise Behavior in Graphene

*Guangyu Xu[1], Carlos M. Torres Jr.[1], Yuegang Zhang[2], Fei Liu[3], Emil B. Song[1], Minsheng Wang[1], Yi Zhou[1], Caifu Zeng[1] and Kang L. Wang[1]*

1. Department of Electrical Engineering, University of California at Los Angeles, Los Angeles, California 90095, USA

2. Molecular Foundry, Lawrence Berkeley National Laboratory, 1 Cyclotron Road, Berkeley, California 94720, USA

3. IBM T. J. Watson Research Center, Yorktown Heights, New York 10598, USA

[*]email: wang@ee.ucla.edu (KW); yzhang5@lbl.gov (YZ)

**ABSTRACT**. Scattering mechanisms in graphene are critical to understanding the limits of signal-to-noise-ratios of unsuspended graphene devices. Here we present the four-probe low frequency noise (1/f) characteristics in back-gated single layer graphene (SLG) and bilayer graphene (BLG) samples. Contrary to the expected noise increase with the resistance, the noise for SLG decreases near the Dirac point, possibly due to the effects of the spatial charge inhomogeneity. For BLG, a similar noise reduction near the Dirac point is observed, but with a different gate dependence of its noise behavior. Some possible reasons for the different noise behavior between SLG and BLG are discussed.

Graphene is a two-dimensional film with intrinsically ultrahigh carrier mobility, showing extraordinary potential for device applications[1-5]. However, most unsuspended graphene sheets are degraded by external perturbations from the environment[3,6,7]. For example, the charged impurity scattering lessens the carrier mobility of substrated graphene devices; the carrier trapping/detrapping near the graphene-substrate interface leads to device variability and contributes to the low frequency (1/f) noise[8,9]. Recent work has shown that the random charged impurities near the graphene-substrate interface creates an inhomogeneous charge distribution along the graphene sheet[10,11]. The spatial-charge inhomogeneity in graphene affects the ideal transport properties and is responsible for several physical anomalies near the Dirac point [12,13]. It would be of fundamental interest to investigate how the presence of spatial charge inhomogeneity influences the 1/f noise behavior in graphene. A detailed study on the 1/f noise behavior of substrated graphene would also help to achieve high-speed carbon-based electronics with high signal-to-noise-ratios.

Multiple groups have conducted research on the 1/f noise behavior on graphene nanostructures. Previous work on graphene nanoribbon (with 30nm width) shows that 1/f noise increases as the resistance increases in single layer graphene nanoribbon (SLR), whereas the noise increases as the resistance decreases in bilayer nanoribbon (BLR), which is attributed to the bandgap-opening effect[8]. For bulk graphene (with micron width), dual-gated device structures have been used to achieve low noise level in single layer graphene (SLG) [14], and to understand the noise correlation with the band structure in bilayer graphene (BLG) [9]. In this work, we present the four-probe low frequency (1/f) noise characteristics in SLG and BLG samples using a back-gated device structure. The back-gated structure helps simplify the interface physics in understanding the carrier-substrate interaction[15,16]. We focus on bulk samples with a four-probe setup to reduce the noise contribution from the edge states (as in nanoribbon)[17,18] and the metal contacts[8,19-21], although the contacts might still affect the electron-hole symmetry in this work[22,23]. For SLG, we find that the noise was reduced either close to or far away from the Dirac point (M-shape); for BLG, we find a similar noise reduction near the Dirac point, but with an increase of the noise away from the Dirac point (V-shape). Our noise data near the Dirac point can be correlated to the spatial-charge inhomogeneity at low carrier density limits; this fact might provide insights to the scattering mechanisms in graphene near the Dirac point.

Graphene sheets were mechanically exfoliated from natural graphite and transferred onto a 300 nm thermally grown $SiO_2$ dielectric film on highly doped Si substrates. Subsequently, they were identified through optical microscopy and Raman spectroscopy, and patterned to form Hall bar and/or multi-probe structures using general e-beam writing processes (inset of Fig. 1a) [23, 24]. The devices are maintained in vacuum environment to avoid contact oxidation and uncontrollable doping effects from the ambiance[24]. Before the measurements, a 20-minute vacuum bakeout (100°C) process is generally applied to partially desorb contaminants (see Supplement) [7]. A total of more than 10 out of 40 back-gated devices were studied in our work, and the data shown in this paper come from 4 SLG and 4 BLG samples.

In the four-probe configuration, an Agilent 4156C was used to apply dc current bias to the device within its linear regime (not shown), and measure its dc conductivity $\sigma$; an Agilent 35670A was used to collect the noise spectra of the fluctuations of the potential difference ($V$) across the graphene samples (see Supplement). Fig. 1a shows the typical room temperature conductivity versus the gate bias (shifted by $V_{Dirac}$, the gate bias at Dirac point) for a SLG sample (SLG1). At each gate bias, the conductivity is averaged by 10 times of measurements at the same time of the noise measurement in order to avoid the hysteresis and ensure the consistency of the data[20]. Fig. 1b shows the room-temperature noise spectra, $S_V$, (after subtracting the background noise [20, 25], see Supplement) of SLG1 at the same gate biases applied in the conductivity measurement (see Fig. 1a). The noise power spectrum density $S_V$ follows a $1/f^\alpha$ behavior with $\alpha$ ranging from 0.85 to 1.12, when biased either close to or away from the Dirac point.

The noise figure $A = f \cdot S_V / V^2$ is commonly used to characterize the noise level for current bias conditions[15]. In this work, the noise is normalized as $A = \frac{1}{Z}\sum_{i=1}^{Z} f_i \cdot S_{Vi} / V^2$, which takes the average over the frequency ranges where the noise spectra follow with $1/f^\alpha$ behavior ($\alpha$ ranging from 0.85 to 1.12). This definition helps reduce the measurement errors of the noise at specific frequencies[8, 20], and rule out other types of noise sources (e.g. thermal noise, ac electricity power noise, etc) [9, 25]. For SLG1, the gate dependence of the normalized noise, $A$, shows an M-shape behavior (see Fig. 1c). The noise curve shows a local minimum at the Dirac point and two local maximum near the Dirac point at both the electron- ($V_g-V_{Dirac}>0$) and hole-conduction ($V_g-V_{Dirac}<0$) sides. The noise reduction observed in SLG near the Dirac point is different from most electronic materials (including SLR), where the noise increases with sample resistance[8].

We measured SLG samples under different conditions to rule out extrinsic possibilities for the M-shape noise behavior. First, this M-shape noise behavior was found to be independent of the types of the initial doping ($V_{Dirac-SLG1}>0$, $V_{Dirac-SLG2}\sim 0$, $V_{Dirac-SLG3}<0$, see Fig. 2a). Second, the measurements have been repeated by sweeping the gate voltage back and forth (not shown), to assure that the noise behavior is independent of the mobile ions close to the graphene-$SiO_2$ interface[26]. Third, the noise level was independent of the current biases (see Supplement), showing that the noise of graphene is from the resistance fluctuations[19,27, 28] and not affected by the current-induced local heating effects[24, 29]. Overall, the M-shape noise behavior is universal in our measurements and suggests to originate from the intrinsic trapping/detrapping processes near the graphene-$SiO_2$ substrate in SLG.

To gain some insights of the M-shape noise behavior, it helps to see where the noise starts to reduce near the Dirac point. The spatial average carrier density could be defined as $<n>=C_g(V_g-V_{Dirac})/e$ [24, 26], when $<n>$ is not less than the density of charged impurities $n_{imp}$ [30]. Fig. 2b shows that the noise maximum generally occurs at $|<n_{noise\max}>|=1\sim 1.6\times 10^{12} cm^{-2}$, a value approximately equal to the estimated $n_{imp}=1\sim 2\times 10^{12} cm^{-2}$ in our SLG samples[31, 32] (Note: the charged impurity density, $n_{imp}$, used in this work is estimated as the theoretical $\sigma_{Dirac}-n_{imp}$ and $\mu-n_{imp}$ relations based on the self-consistent approximation, see Ref 31 and 32). It is accepted that the transport properties of SLG near the Dirac point (i.e.$|<n>| < n_{imp}$) are dominated by the spatial charge inhomogeneity, when the landscape of SLG is broken into puddles of electrons and holes[30, 33]. Our data show that the noise of SLG starts to reduce when $|<n>| < n_{imp}$, which suggest that the noise behavior near the Dirac point correlates with the spatial charge inhomogeneity. The value of the noise maximum, $A_{max}$, is almost equal for both the electron- and hole-conduction sides ($10^{-8} \sim 10^{-7}$), while some differences might be from the electron-hole asymmetry induced by the doping effect of metal contacts[23].

Up to now, a noise model of graphene materials considering the spatial charge inhomogeneity is still lacking[8, 9]. Admitting the complexity of this problem, here we try to use a qualitative approach to give some phenomenological explanations. Some assumptions are made to simplify the discussion:

1) Close to the Dirac point, electrons and holes transport along the puddles of electrons and holes, respectively[26, 30], hence both types of carriers would contribute to the noise. By neglecting the electron-hole interaction, we assume these two types of carriers independently interact with the active traps (see Supplement) in their puddle regions, respectively.

2) Previous work shows that two types of puddle regions exist in SLG near the Dirac point[26, 32, 33]: I) wide regions (over the whole graphene sheet) with almost uniform low carrier densities[26, 32]; II) narrow regions (typical size ~ 10nm) with much higher carrier densities. We neglect the contribution of the noise from type-II regions since they are highly conductive (better screening to active traps) and only occupy a small portion (< 20%) [32, 33]. Hence, the noise mainly comes from the type-I regions, where the carrier density is almost uniform ($n_{electron}^{TypeI} \sim const, n_{hole}^{TypeI} \sim const$).

3) Hooge's empirical equation $A = f \cdot S_V / V^2 = \alpha_H / N$ is well-defined for homogenous materials[27, 34]; this equation generally needs to be modified for inhomogeneous materials by weighting each local area differently[21, 27]. Based on 1) and 2), however, we see the total noise mainly comes from two independent type-I regions for both electrons and holes, respectively. Since type-I regions are nearly homogenous, the total noise could be estimated as $A = A_{electron} + A_{hole} \sim \frac{\alpha_H}{N_{electron}^{TypeI}} + \frac{\alpha_H}{N_{hole}^{TypeI}} \sim \frac{\alpha_H}{n_{electron}^{TypeI} D_{electron}^{TypeI}} + \frac{\alpha_H}{n_{hole}^{TypeI} D_{hole}^{TypeI}}$ ($A_{electron(hole)}$ is the noise from electrons(holes); $N^{TypeI}_{electron(hole)}$ and $D^{TypeI}_{electron(hole)}$ are the total number of electrons (holes) and the area of type-I regions for electrons (holes), respectively) [8, 9, 14]. The electron-electron (hole-hole) interaction is neglected, because Hooge's law implies that each carrier is considered to interact with active traps independently[27, 34]

In the following, we limit the discussion to the noise behavior at the Dirac point and the electron-conduction side ($V_g - V_{Dirac} > 0$), since the hole-conduction side ($V_g - V_{Dirac} < 0$) could be explained similarly.

1) The noise at the Dirac point

At the Dirac point, charge neutrality is satisfied ($N^{TypeI}_{electron} \sim N^{TypeI}_{hole}$). Due to the existence of the spatial charge inhomogeneity, the total noise is given as $A \sim \frac{2\alpha_H}{N^{TypeI}_{electron}} \sim \frac{2\alpha_H}{n_{rms} D^{TypeI}_{electron}}$ (here we define $n_{rms}$ as the rms of spatial density fluctuations at the Dirac point, and $|n^{TypeI}_{electron}| \sim |n^{TypeI}_{hole}| \sim n_{rms}$ at the Dirac point [32, 33]). Thus, the noise at the Dirac point is determined from the carrier density fluctuations caused by spatial charge inhomogeneity in SLG.

2) Noise behavior before the noise maximum

Between the Dirac point and the noise maximum, one type of carriers (i.e. majority) would dominate the other (i.e. minority). As the gate bias increases ($V_g - V_{Dirac} > 0$): a) $n^{TypeI}_{electron(hole)}$ increases (decreases) (i.e. $n^{TypeI}_{hole} < n^{TypeI}_{electron}$), and $n^{TypeI}_{electron(hole)} \propto V_g - V_{Dirac}$ [24, 26] (type-I regions are nearly homogeneous). Thus, $\Delta(n^{TypeI}_{hole}) \sim \Delta(n^{TypeI}_{electron})$. b) $D^{TypeI}_{electron(hole)}$ increases (decreases) (i.e. $D^{TypeI}_{hole} < D^{TypeI}_{electron}$), and $\Delta(D^{TypeI}_{hole}) \sim \Delta(D^{TypeI}_{electron})$. c) The noise $A_{electron(hole)} \sim \frac{\alpha_H}{N^{TypeI}_{electron(hole)}} \sim \frac{\alpha_H}{n^{TypeI}_{electron(hole)} \cdot D^{TypeI}_{electron(hole)}}$ decreases (increases). Thus, $A_{hole} > A_{electron}$. The overall effect is $\Delta(A_{hole}) \sim A_{hole} \times \left[ \frac{\Delta(n^{TypeI}_{hole})}{n^{TypeI}_{hole}} + \frac{\Delta(D^{TypeI}_{hole})}{D^{TypeI}_{hole}} \right] > A_{electron} \times \left[ \frac{\Delta(n^{TypeI}_{electron})}{n^{TypeI}_{electron}} + \frac{\Delta(D^{TypeI}_{electron})}{D^{TypeI}_{electron}} \right] \sim \Delta(A_{electron})$, an increase of noise from holes greater than a decrease of noise from electrons. Therefore, the total noise increases ($\Delta A = \Delta(A_{hole}) - \Delta(A_{electron}) > 0$) as we move away from the Dirac point before reaching the noise maximum.

3) Noise behavior beyond the noise maximum

As the gate bias keeps increasing, $D_{hole}^{TypeI}$ continues shrinking and the hole-puddle region becomes sparse and cannot form a path for hole-conduction[26,32,33]. When those isolated hole islands stop contributing to the noise (even though they have small $n_{hole}^{TypeI}$ and $D_{hole}^{TypeI}$), the total noise only comes from the electrons as $A \sim A_{electron} \sim \frac{\alpha_H}{N_{electron}^{TypeI}}$. Hence, the total noise will decrease as the gate bias further increases, because more electrons effectively screen the active traps.

Based on 2) and 3), we suggest the noise maximum point occurs when the holes (minority) stop dominating the noise behavior (become isolated and not conducting). Under this physical picture, we estimate the noise maximum point as follows: at the Dirac point $|n_{hole}^{TypeI}| \sim n_{rms}$, whereas at the noise maximum $|n_{hole}^{TypeI}|$ becomes negligible. In order to modulate $|n_{hole}^{TypeI}|$ from $n_{rms}$ to near zero, a gate bias that is equivalent to the spatial average density $\sim n_{rms}$ is required ($n_{hole}^{TypeI} \propto V_g - V_{Dirac}$). Thus, the noise maximum occurs near $|<n_{noise\ max}>| \sim n_{rms}$. This estimate is consistent with the experimental data in Fig. 2b ($|<n_{noise\ max}>| \sim n_{imp}$), since theoretical works show that $n_{rms} \sim n_{imp}$ for typical $n_{imp} \sim 10^{11} - 10^{12} cm^{-2}$ [31,32].

To further understand the noise mechanism, we measured another SLG sample (SLG4) under temperatures from 180K to 90K (see Fig. 2c), where the noise spectra all follow the 1/f behavior. The noise at all gate biases decreases when temperature decreases, possibly because some active traps become frozen ($kT$) and stop contributing to the noise[27, 28]. The weak temperature-dependence of the dc conductivity (see inset) confirms that the transport is limited by charged impurity scattering [24, 33]. Similarly, the M-shape noise behavior is shown for all temperatures, and the noise maximum points (for both the electron- and hole-conduction sides) do not change with temperature. The data suggest that the distribution of spatial charge inhomogeneity is almost unchanged ($n_{imp}$, $n_{rms}$) with the temperature as predicted [31, 32]. It is noted that the region of the noise reduction near the Dirac point (the dip in M-shape) is narrower (smaller $|<n_{noise\ max}>|$) than those in SLG1-SLG3. This can be attributed to a smaller spatial charge inhomogeneity in SLG4. ($n_{imp}/n_{rms} \sim 7 \times 10^{11} cm^{-2}$, estimated carrier mobility (Ref. 24) $\mu_e, \mu_h \sim 4000 cm^2/(V \cdot s)$ whereas $\mu_e, \mu_h < 3000 cm^2/(V \cdot s)$ for SLG1-SLG3).

So far, we attributed the M-shape noise behavior of SLG to the spatial charge inhomogeneity. For our SLG samples, the M-shape noise behavior is consistently observed. However, we do not deny that some high quality SLG samples ($n_{imp}, n_{rms} \to 0$) might not show the noise reduction near the Dirac point ($|<n_{noise\ max}>| \to 0$). The fact that our noise spectra do not deviate from 1/f behavior at low temperature suggests the existence of multiple active traps (>>1) in our SLGs[19, 21], where the Hooge parameter $\alpha_H$ is estimated to be $10^{-3} \sim 10^{-2}$ away from $V_{Dirac}$ for both electron- and hole-conduction sides. Moreover, detailed noise modeling considering the carrier-carrier interactions (e-e, h-h, e-h interactions) and the noise contribution from the small type-II regions is needed to further understand the noise behavior quantitatively.

We also examine the noise behavior of BLG samples. Three back-gated BLG samples are measured ($V_{Dirac-BLG3} < V_{Dirac-BLG1} < V_{Dirac-BLG2} < 0$) at room temperature; they all exhibit a V-shape noise behavior (see Fig. 3a), showing a similar noise reduction near the Dirac point to that of SLG. Temperature-dependent measurement (BLG4, from 300K to 77K) exhibits results qualitatively similar to SLG (see Fig. 3b): the transport shows weak temperature-dependence; the V-shape noise behavior is observed for all temperatures, where the 1/f noise spectra remain and $\alpha_H$ ranges in the same order of those in SLGs.

A recent work observed the V-shape noise behavior in dual-gated BLG samples as well[9], which has been attributed to the bandgap-opening in the BLG band structure. While this might be the reason, the gate-induced bandgap in back-gated BLG still lacks experimental evidence from electrical transport measurement[35, 36]

Admitting that the effect of the bandgap-opening on the noise behavior of BLG may be possible, here we try to consider another scenario. In a gapless situation, the transport of BLG near Dirac point has been suggested to resemble that of SLG where the spatial charge inhomogeneity breaks the density landscape into puddles of electrons and holes[37, 38]. Indeed, a recent STM experiment shows the existence of spatial charge inhomogeneity in back-gated BLG at the Dirac point[39]. Hence, we propose that the noise behavior of BLG near the Dirac point could also be correlated with the spatial charge inhomogeneity: At the Dirac point, the noise is determined by $|n_{electron}^{TypeI}| \sim |n_{hole}^{TypeI}| \sim n_{rms}$, and away from Dirac point, the noise increases possibly because of the noise increase from the minority carriers (e.g. holes for $V_g\text{-}V_{Dirac} > 0$), $\Delta A = \Delta(A_{minority}) - \Delta(A_{majority}) > 0$. The noise in BLG, however, does not decrease at higher biases as the case in SLG, showing a V-shape instead of an M-shape. The reason may come from the large spatial charge inhomogeneity (even in clean samples) of BLG: due to their different dispersion relationships, BLG has been predicted to have a larger $n_{rms}$ than that in SLG under the same disorder level ($n_{imp}$) [37, 38]. Hence, BLG would qualitatively have a wider gate range where the noise is contributed from both electrons and holes, and increases with the gate bias (refer to the noise behavior between the Dirac point and the noise maximum in SLG). Overall, the V-shape noise behavior of BLG could be from the bandgap-opening effect and/or spatial charge inhomogeneity.

In conclusion, we present the low frequency noise behavior in back-gated SLG (M-shape) and BLG (V-shape) samples, both with weak temperature-dependence (down to 77K). Using a qualitative approach, the noise behavior near the Dirac point suggests relevance to the spatial charge inhomogeneity in SLG and BLG samples. A quantitative noise model of graphene is necessary to deepen the understanding. Fundamentally, low frequency noise describes the trapping/detrapping processes near the graphene-$SiO_2$ interface, while more efforts are needed to gain further insights on the scattering mechanisms of graphene. Moreover, this work may help predict, control and improve the signal-to-noise-ratio of substrated graphene devices and suppress the phase distortion for high frequency applications.

**ACKNOWLEDGMENT** The authors gratefully acknowledge the discussions from F. Miao, X. Zhang and F. X. Xiu, and experimental help from S. Aloni, T. Kuykendall, Z. J. Xu and J. W. Bai. We thank E. Rossi, E. H. Hwang and S. Adam from S. Das Sarma's group for theoretical discussions. This work was in part supported by MARCO Focus Center on Functional Engineered Nano Architectonics (FENA), monitored by Dr. Betsy Weitzman. The work at the Molecular Foundry was supported by the Office of Science, Office of Basic Energy Sciences, of the U.S. Department of Energy under Contract No. DE-AC02-05CH11231.

**REFERENCES**

1. Novoselov. K.S. et. al. Electric field effect in atomically thin carbon films. *Science*, **306**, 666-669 (2004)

2. Morozov, S. V. et. al. Giant intrinsic carrier mobilities in graphene and its bilayer. *Phys. Rev. Lett.* **100**, 016602 (2008)

3. Du, X. et. al. Approaching ballistic transport in suspended graphene. *Nature Nanotech*. **3**, 491-495 (2008)

4. Hong, X. et. al. High-mobility few-layer graphene field effect transistors fabricated on epitaxial ferroelectric gate oxide. *Phys. Rev. Lett.* **102**, 136808 (2009)

5. Lin, Y.M. et. al. Operation of graphene transistors at gigahertz frequencies. *Nano Lett.*, **9**, 422-426 (2009)

6. Chen. J. H. et. al. Intrinsic and extrinsic performance limits of graphene devices on $SiO_2$. *Nature Nanotech*, **3**, 206-209 (2008)

7. Chen. J. H. et. al. Charged-impurity scattering in graphene. *Nature Phys*, **4**, 377-381 (2008)

8. Lin, Y. M. and Avouris, P. Strong suppression of electrical noise in bilayer graphene nanodevices. *Nano Lett.*, **8**, 2119-2125 (2008)

9. Pal, A. N. and Ghosh, A. Resistance noise in electrically biased bilayer graphene. *Phys. Rev. Lett.* **102**, 126805 (2009)

10. Martin, J. et. al. Observation of electron-hole puddles in graphene using a scanning single-electron transistor. *Nature Phys*, **4**, 144-148 (2008)

11. Zhang, Y.B. et. al. Origin of spatial charge inhomogeneity in graphene. *Nature Phys*. **5**, 722-726 (2009)

12. Heersche, H. B. et. al. Bipolar supercurrent in graphene. *Nature*, **446**, 56-59 (2007)

13. Wei, P. et. al. Anomalous thermoelectric transport of Dirac particles in graphene. *Phys. Rev. Lett.* **102**, 166808 (2009)

14. Gang. L. et. al. Low-frequency electronic noise in the double-gate single-layer graphene transistors, *Appl. Phys. Lett.* **95**, 033103 (2009)

15. Lin, Y. M. et. al. Low-frequency current fluctuations in individual semiconducting single-wall carbon nanotube. *Nano Lett.*, **6**, 930-936 (2006)

16. Jang, C. et. al. Tuning the effective fine structure constant in graphene: Opposing effects of dielectric screening on short- and long-range potential scattering. *Phys. Rev. Lett.* **101**, 146805 (2008)

17. Evaldsson, M. et.al. Edge-disorder-induced Anderson localization and conduction gap in graphene nanoribbons. *Phys. Rev. B*. **78**, 161407 (R) (2008)

18. Barone, V. et. al. Electronic structure and stability of semiconducting graphene nanoribbons. *Nano*

*Lett.*, **6**, 2748-2754 (2006)

19. Tobias, D. et.al. Origins of 1/f noise in individual semiconducting carbon nanotube field-effect transistors. Phys. Rev. B. 77, 033407 (2008)

20. Xu, G. et. al. Low-frequency noise in top-gated ambipolar carbon nanotube field effect transistors. *Appl. Phys. Lett.* **92**, 223114 (2008)

21. Kolek, A. et. al. Low-frequency 1/f noise of $RuO_2$-glass thick resistive films. *J. Appl. Phys.* **102**, 103718 (2007).

22. Lee, E. J. H. et. al. Contact and edge effects in graphene devices. *Nature Nanotech.* **3**, 486-490(2008)

23. Han, W. Electron-hole asymmetry of spin injection and transport in single-layer graphene. *Phys. Rev. Lett.* **102**, 137205 (2009)

24. Tan, Y. W. et. al. Measurement of scattering rate and minimum conductivity in graphene. *Phys. Rev. Lett.* **99**, 246803 (2007)

25. Snow. E. S. et. al. 1/f noise in single-walled carbon nanotube devices. *Appl. Phys. Lett.* **85**, 4172-4174 (2004)

26. Rossi, E. and Sarma, S. D. Ground state of graphene in the presence of random charged impurities. *Phys. Rev. Lett.* **101**, 166803 (2008)

27. Van Der Zeil, A. Unified presentation of l/f noise in electronic devices: fundamental l/f noise sources. *Proc. of IEEE*, **76**, **No. 3**, 233-258 (1988)

28. Huang, K.K. et. al. A unified model for the flicker noise in metal-oxide-semiconductor field-effect transistors. *IEEE Trans. Electron Devices,* **37**, 654-665 (1990)

29. Barreiro, A. et. al. Transport properties of graphene in the high-current limit. *Phys. Rev. Lett.* **103**,

076601 (2009)

30. Hwang E. H. et. al. Carrier transport in two-dimensional graphene layers. *Phys. Rev. Lett.* **98**, 186806 (2007)

31. Adam, S. et. al. A self-consistent theory for graphene transport. *Proc. Natl Acad Sci. USA.* **104**, 18392-18397 (2007)

32. Rossi, E. et. al. Effective medium theory for disordered two-dimensional graphene. *Phys. Rev. B.* **79**, 245423 (2009)

33. Adam, S. et. al. Theory of charged impurity scattering in two-dimensional graphene. *Solid State Commun.* **149**, 1072-1079 (2009)

34. Hooge, F.N. l/f noise sources. *IEEE Trans. Electron Devices,* **41**, 1926-1935 (1994)

35. Zhang, Y.B. et. al. Direct observation of a widely tunable bandgap in bilayer graphene. *Nature.* **459**, 820-823(2009)

36. Oostinga, J. B. et. al. Gate-induced insulating state in bilayer graphene devices. *Nature Mater.* **7**, 151-157 (2008)

37. Adam, S. and Sarma, S. D. Boltzmann transport and residual conductivity in bilayer graphene. *Phys. Rev. B.* **77**, 115436 (2008)

38. Sarma, S. D. et. al. Theory of carrier transport in bilayer graphene. *arXiv. 0912.0403v1*, Dec. 2$^{nd}$ (2009)

39. Danshpande, A. et. al. Mapping the Dirac point in gated bilayer graphene. *Appl. Phys. Lett.* **95**, 243502 (2009)

**FIGURE CAPTIONS**

**Figure 1 Characterization of the bulk single layer graphene (SLG) using dc conductivity and noise measurement. a.** Room temperature dc conductivity versus gate bias ($V_g$-$V_{Dirac}$) for SLG1. The inset shows the optical pictures of the back-gated Hall bar and multi-probe structures. **b.** Room temperature low frequency noise spectrum of SLG1 with gate biases varying from -50V to 100V (step = 5V). The background noise measured at zero current bias has been subtracted from the raw noise data. Four-probe noise spectra ($S_V$) follow $1/f^\alpha$ behavior with $\alpha$ ranging from 0.85 to 1.12 by biasing $V_g$ both near to and far away from the Dirac point. **c.** Noise (A) versus gate bias ($V_g$-$V_{Dirac}$) for SLG1. An M-shape noise behavior is observed: a noise minimum occurs at the Dirac point, and the two noise maximum points (see the dashed lines) occur on both the electron and hole sides. Neither field-induced nor quantum-confined bandgap could contribute to this noise behavior near the Dirac point[8, 26, 27]

**Figure 2 Noise data of single layer graphene (SLG). a.** Room temperature noise (A) versus gate bias ($V_g$-$V_{Dirac}$) for SLG1, SLG2 and SLG3. All SLG samples show an M-shape noise behavior, which is independent of the types of the initial doping ($V_{Dirac-SLG1}$ >0, $V_{Dirac-SLG2}$ ~0, $V_{Dirac-SLG3}$ <0). The noise behavior is repeatable by changing the direction of the injection current through the samples. **b.** Spatial average carrier density of the noise maximum ($|<n_{noise\ max}>|$) versus the value of noise maximum ($A_{max}$), for both the electron-conduction (solid circles, $V_g$-$V_{Dirac}$>0) and hole-conduction (hollow circles, $V_g$-$V_{Dirac}$<0) sides. The noise maximum points at both sides were shown at $|<n_{noise\ max}>|=1\sim1.6\times10^{12} cm^{-2}$. It is estimated that $n_{imp}=1\sim2\times10^{12}cm^{-2}$ ($r_s=0.8, d=1nm$) in our SLG samples using the theoretical methods[31, 32]. **c.** Temperature dependence of the noise behavior for SLG4. The M-shape noise behavior does not change with the temperature down to 90K. The noise level is shown to monotonously decrease with the temperature. The inset shows the weak temperature-dependence of dc conductivity versus the gate bias ($V_g$-$V_{Dirac}$) for SLG4. The dashed vertical lines show that the noise maximum is independent of the temperature.

**Figure 3 Noise data of bilayer graphene (BLG). a.** Room temperature noise (A) versus gate bias ($V_g$-$V_{Dirac}$) for BLG1, BLG2 and BLG3. All BLG samples show a V-shape noise behavior ($V_{Dirac\text{-}BLG3}$ < $V_{Dirac\text{-}BLG1}$ < $V_{Dirac\text{-}BLG2}$ < 0). The noise measurements have been repeated to rule out the effects from mobile ions and local heating. The V-shape noise behavior is shown to be universal in our BLG samples. **b.** Temperature dependence of the noise behavior for BLG4. The V-shape noise behavior does not change with the temperature from 300K to 77K. Similar to SLG, the noise level is shown to monotonously decrease with the temperature, possibly because some active traps are frozen. The inset shows the weak temperature-dependence of dc conductivity versus the gate bias ($V_g$-$V_{Dirac}$) for BLG4.

**Figure 1**
**Xu et. al.**

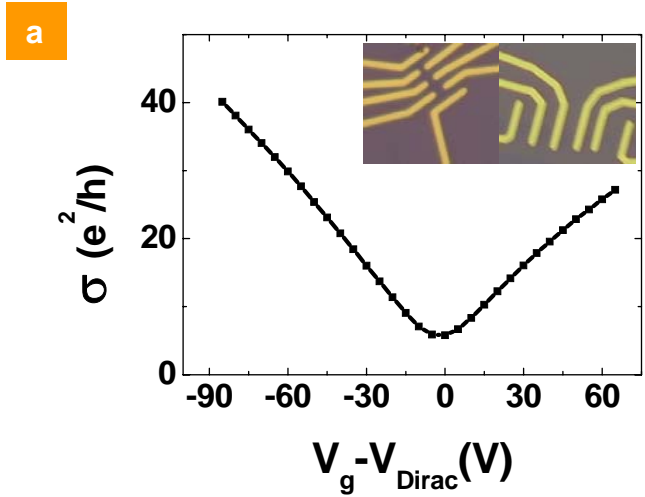
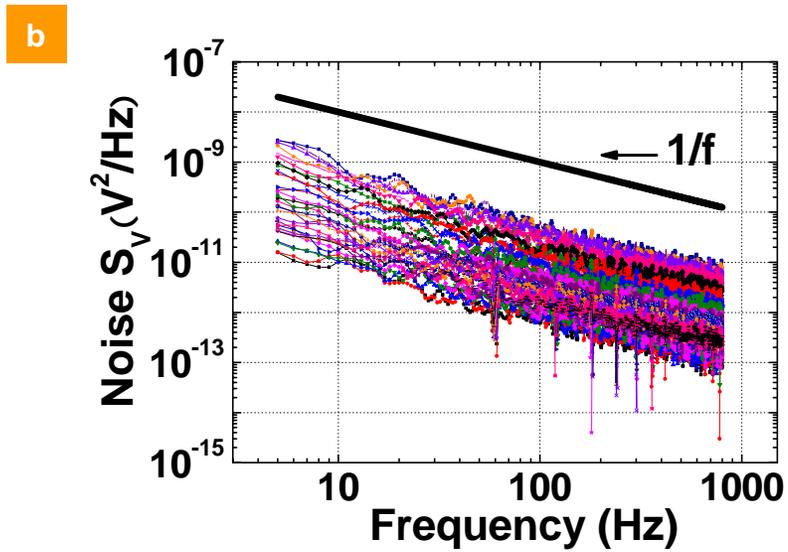
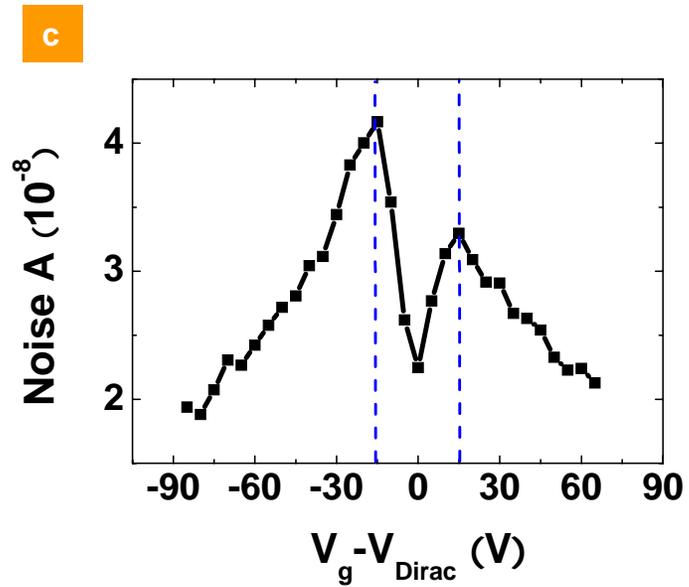

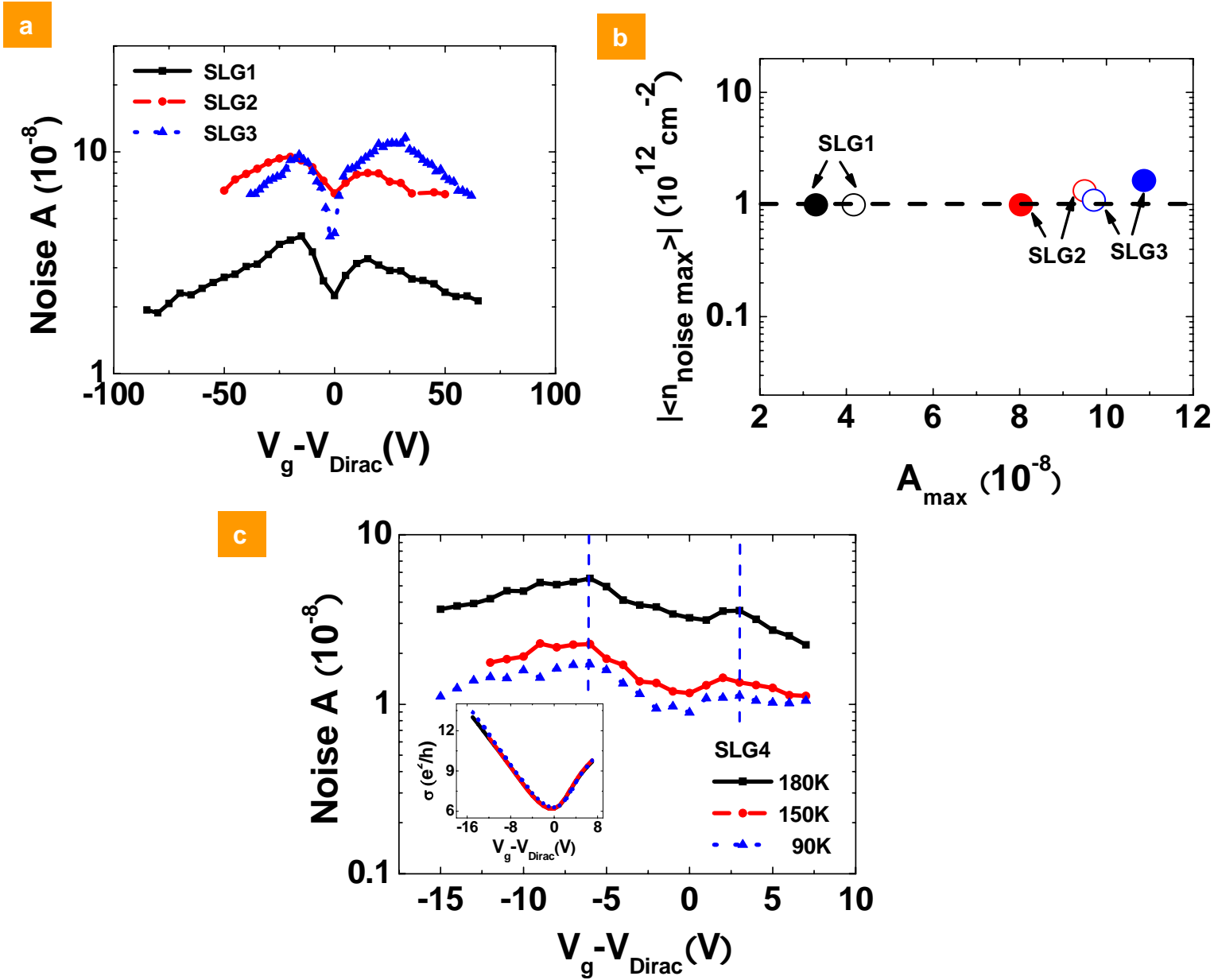

**Figure 2**
**Xu et. al.**

**Figure 3**
**Xu et. al.**

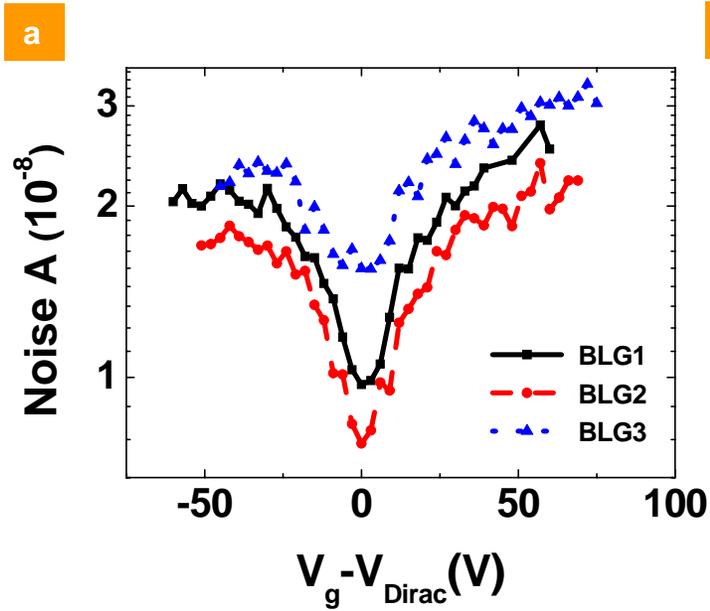
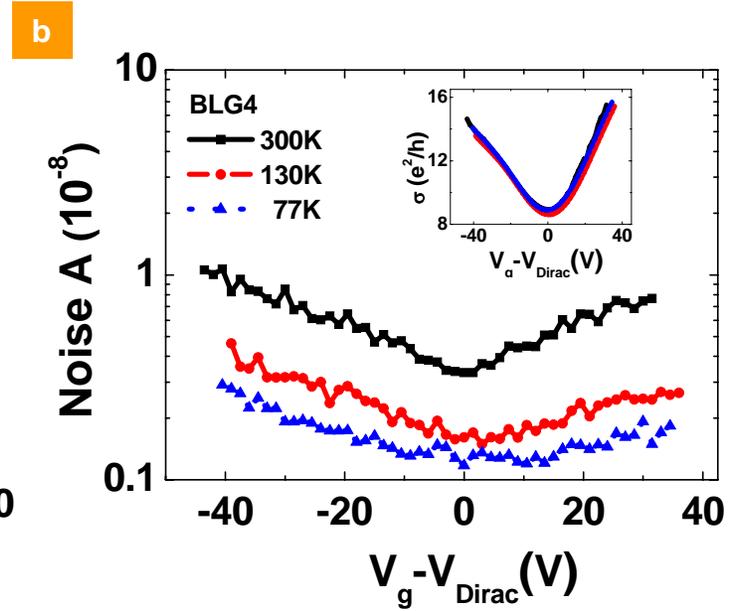

# Effect of Spatial Charge Inhomogeneity on 1/f Noise Behavior in Graphene

G. Xu, C. M. Torres Jr., Y. Zhang, F. Liu, E. B. Song, M. Wang, Y. Zhou, C. Zeng and K. L. Wang

**Supplementary Materials**

I.       Methods

**<u>DEVICE PREPARATION</u>**

In this work, a thermally grown $SiO_2$ layer on a highly doped Si wafer is used as the bottom gate dielectric for the graphene devices. The thickness of the $SiO_2$ layer varies from 301 nm to 326 nm for different device batches. After the surface of $SiO_2$ is cleaned by acetone, isopropanol and oxygen plasma, graphene sheets are exfoliated and placed onto the $SiO_2$ substrate from the natural graphite flakes. The graphene sheets are characterized by optical microscopy and Raman spectroscopy. Figure S1a and S1b show the SEM pictures for the typical SLG and BLG, respectively. Figure S1c and Figure S1d shows the Raman spectroscopy for a typical SLG and BLG, respectively.

Then, a MMA/PMMA based dual-layer spin coating is applied followed by a 2 minute $150^{\circ}C$ baking. After an e-beam lithography step, a Titanium/Gold metal layer is evaporated to serve as the electrical contacts with a 7 nm/80 nm thickness, respectively. The devices are maintained in vacuum environment to avoid contact oxidation and uncontrollable doping effects from the ambiance. A 20 minute $100^{\circ}C$ vacuum bakeout process is generally applied to partially desorb contaminants.

The graphene sheet dimensions used in this work is listed in Table I. [S1]. All devices show good ohmic contact with an injected current up to 2mA. With the gate biases ranging from -100V to 100V, the gate leakage current is generally below 4nA at room temperature.

**NOISE MEASUREMENT**

The noise measurement is performed in a Janis 500 four-arm probe station, pumped down to a vacuum of $10^{-6}$ torr at room temperature. A copper sample stage is used as the back-gate and contacted by the cold finger. The graphene devices are fixed onto the stage through the double-sided copper tapes. The gate bias is added through the copper tape to the highly doped Si, where the series resistance is generally below 1kΩ. Figure S2 shows the schematics of the four probe noise measurement setup. An Agilent 4156C is used to apply dc current bias ($I_{inj}$) to the device and to measure its dc conductivity. A SR560 is calibrated to amplify the voltage difference across the graphene sample ($V_{OUT}$), and an Agilent 35670A is used to measure the low frequency noise spectrum typically from f=1Hz to f=800Hz. The noise data are averaged 20 times from the fast Fourier transform (FFT) of the time-domain sampling data, and subtracted by the background noise, which was measured at zero current bias. The conductivity is averaged by 10 times of measurements via the 4156C at the same time of the noise measurement in order to ensure the consistency of the data. The temperature is controlled by a SI9700, and the vapor flow is adjusted by the needle valve of the transfer line. During the low temperature measurement, the temperature fluctuation is less than 4mK and the vacuum change is less than 3%.

## II.    Definition of the noise figure

a) The noise figure, $A$, could be further normalized as $A \sim \frac{1}{A_G}$, where $A_G$ is the sheet area of the graphene samples[S2]. Up to now, there still exists a debate over the dependence of noise on the device dimension in the nano-scale system[S3]. In this work, the effect of spatial charge inhomogeneity to the noise level makes the physical meaning of the dimension-scaled noise complicated. Hence, the noise figure ($A$) without further dimension scaling is used in this work. All conclusions made in this work about the noise behaviors (for both SLG and BLG) are still valid if the noise is scaled by $A_G$.

b) The noise in Ref. S2 is defined as the integration over the whole frequency window, which reduces the measurement errors of the noise at specific frequencies[S1]. However, the original definition in Ref. S2 is frequency-dependent and may not rule out the noise spectra which deviate from the typical 1/f behavior. Hence, the noise in this work is given by $A = \frac{1}{Z}\sum_{i=1}^{Z} f_i \cdot S_{V_i}/V^2$, which takes the average over the frequency ranges where the noise follows with 1/f behavior (α normally ranges from 0.85 to 1.12). This method rules out other types of noise sources (e.g. thermal noise, ac electricity power noise, etc).

### III. Noise independence on the current biases

In this work, the graphene devices are current-biased ($I_{inj}$) in the linear regime for four-probe measurement. Figure S3a and S3b show typical room temperature noise versus $I_{inj}$ (left scale) under typical gate biases for SLG2 and BLG1, respectively. The noise is almost independent of the current biases, which suggests that the local heating effects are insignificant[S4, S5]. On the right scale of Fig. 3a and Fig. 3b, the linear $V_{OUT} - I_{inj}$ curve shows that our devices are biased in the linear regime. The measurement has been repeated for all the other devices used in this work.

### IV. Discussion of the active traps and its relation with charged impurities

The concepts of active traps and charged impurities are different: 1) the traps with their energy near the Fermi energy of the graphene ($\sim kT$) are active in the trapping/detrapping processes and contribute to the 1/f noise; 2) the charged impurities result in the spatial charge inhomogeneity, which defines the puddle regimes. The carriers interact with the active traps in the puddle regimes; this is believed to be the origin of 1/f noise in our graphene samples.

# References

S1. Lin, Y. M. and Avouris, P. Strong suppression of electrical noise in bilayer graphene nanodevices. *Nano Lett.*, **8**, 2119-2125 (2008)

S2. Pal, A. N. and Ghosh, A. Resistance noise in electrically biased bilayer graphene. *Phys. Rev. Lett.* **102**, 126805 (2009)

S3. Snow. E. S. et. al. 1/f noise in single-walled carbon nanotube devices. *Appl. Phys. Lett.* **85**, 4172-4174 (2004)

S4. Tan, Y. W. et. al. Measurement of scattering rate and minimum conductivity in graphene. *Phys. Rev. Lett.* **99**, 246803 (2007)

S5. Barreiro, A. et. al. Transport properties of graphene in the high-current limit. *Phys. Rev. Lett.* **103**, 076601 (2009)

**Table I.** Sheet dimensions of the SLG and BLG samples used in this work together with the data of SLR and BLR from Ref. S1. Sheet dimension was defined as shown in Fig. S2.

TABLE I. Sheet Dimension of the Graphene Samples

| Sample Name | W(μm) | L(μm) | Sample Name | W(μm) | L(μm) |
|---|---|---|---|---|---|
| SLG1 | 6.3 | 2.1 | BLG1 | 4.5 | 2.4 |
| SLG2 | 3.7 | 1.5 | BLG2 | 10.9 | 5.5 |
| SLG3 | 14.4 | 2.1 | BLG3 | 4.8 | 2.1 |
| SLG4 | 2.6 | 2.8 | BLG4 | 4.3 | 1.9 |
| SLR[S1] | 0.03 | 1.7 | BLR[S1] | 0.03 | 2.8 |

**Supplementary Figures**

**Figure S1. Characterization of graphene sheets a.** SEM image (inset) and Raman spectroscopy of a typical SLG. The 2D peak is perfectly fitted with a single Lorentzian shape. **b.** SEM image (inset) and Raman spectroscopy of a typical BLG. The 2D peak could be fitted with a combination of four Lorentzian peaks caused by the band degeneracy of BLG.

**Figure S2. Four-probe configuration of 1/f noise measurement.**

**Figure S3. Noise independence on current biases. a.** Noise versus current bias $I_{inj}$ (left scale) for SLG2 at $V_g - V_{Dirac} = 0V$ under room temperature. The $V_{OUT} - I_{inj}$ curve (right scale) shows the device is biased in linear regime. **b.** Noise versus current bias $I_{inj}$ for BLG1 at $V_g - V_{Dirac} = 20V$ under room temperature. The $V_{OUT} - I_{inj}$ curve (right scale) shows the device is biased in linear regime. The noise is almost independent of the current biases, suggesting that the local heating effects are insignificant.

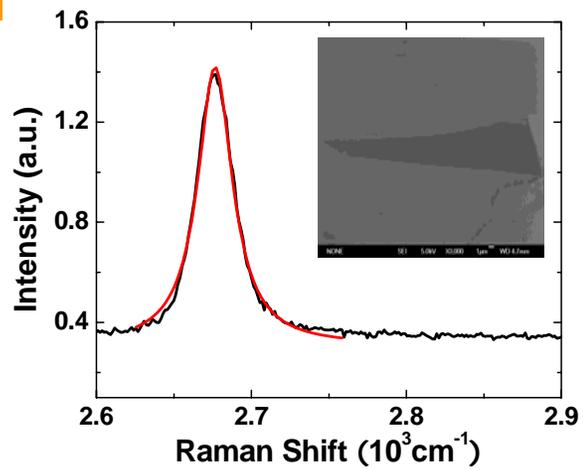 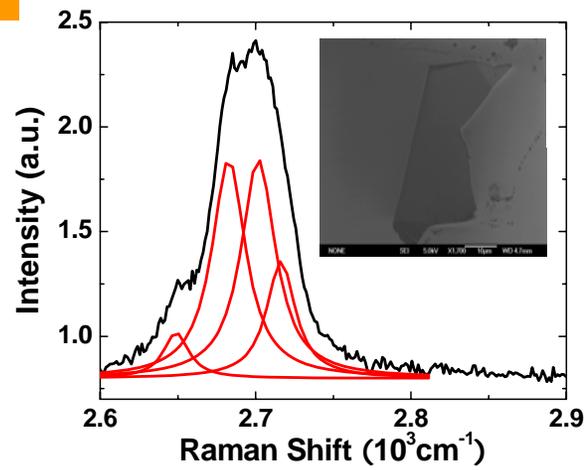

**Figure S2**
**Xu et. al.**

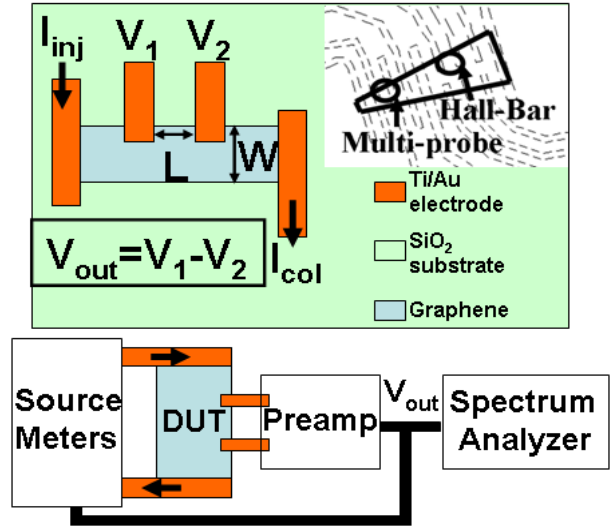

# Figure S3
# Xu et. al.

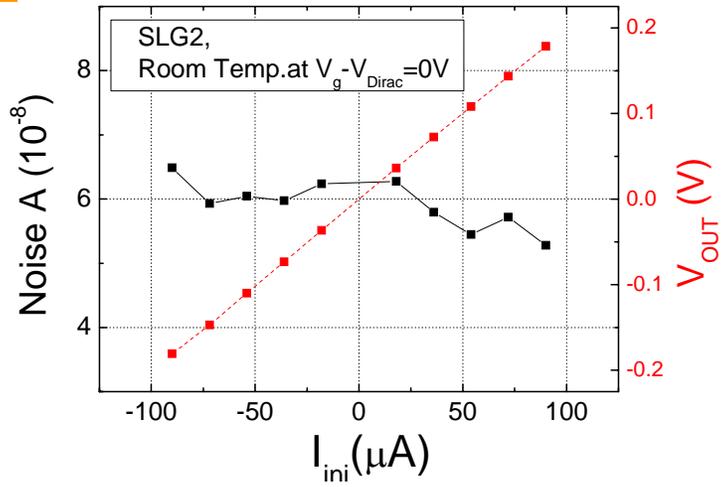 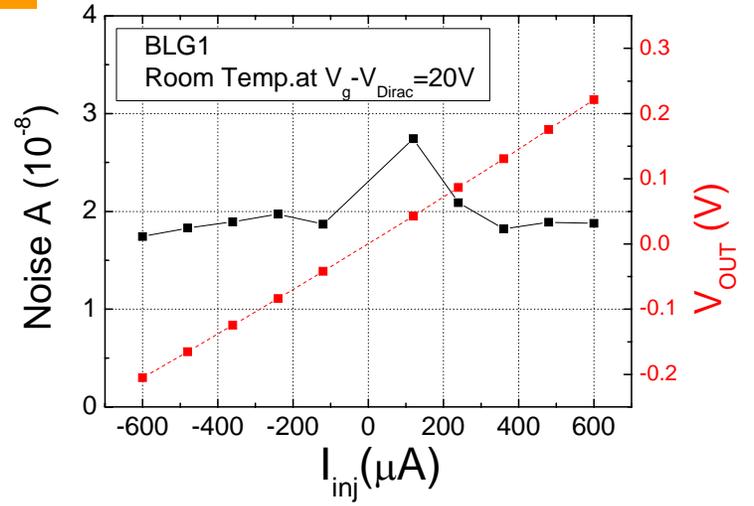